\documentclass[11pt]{article}

\usepackage[preprint]{acl}

\usepackage{times}
\usepackage{latexsym}

\usepackage[T1]{fontenc}
\usepackage[utf8]{inputenc}
\usepackage{microtype}
\usepackage{inconsolata}
\usepackage{graphicx}
\usepackage{amsmath}
\usepackage[table]{xcolor}
\usepackage{subcaption}
\usepackage{booktabs}
\usepackage[table]{xcolor}
\usepackage{makecell}

\title{Few-Shot Contrastive Adaptation for Audio Abuse Detection in Low-Resource Indic Languages}

\newcounter{RZNumberOfComments}
\stepcounter{RZNumberOfComments}

\author{Aditya Narayan Sankaran, Reza Farahbakhsh, Noel Crespi\\
        SAMOVAR, Télécom SudParis\\Institut Polytechnique de Paris
\\91120 Palaiseau, France\\
\texttt{\href{mailto:aditya.sankaran@ip-paris.fr}{aditya.sankaran@ip-paris.fr}}
}

\usepackage{lipsum}
\begin{document}

\maketitle

\begin{abstract}
Abusive and hateful speech is increasingly spoken rather than written, surfacing in voice notes, calls, and short-form videos. Most detection systems still transcribe speech to text before classifying it, but transcription is unreliable for languages lacking strong speech recognisers, and it discards the tone and emotion that often carry the abuse itself. This paper examines whether abusive speech can instead be detected directly from audio, using CLAP, a model that learns a shared representation of sound and language, evaluated across ten Indic languages in the ADIMA dataset. A lightweight classifier trained on CLAP's existing audio representations, without adapting the model itself, comes within one to three points of a fully supervised system, and far outperforms prompting with no labelled examples at all. Further adaptation with a handful of labelled examples per language yields little extra benefit, varying unpredictably across languages. CLAP-based audio representations thus already offer a strong, inexpensive foundation for detecting abusive speech across languages, lowering the labelled data needed in practice.
\end{abstract}

\section{Introduction}

Spoken content is now a primary mode of interaction across short-form video, voice notes, and live audio platforms, making abusive speech in audio an increasingly important moderation problem. Audio-based abuse can be more impulsive and emotionally charged than text~\cite{sakhare2023deep}, yet direct audio abuse detection remains comparatively underexplored~\cite{sharon2022mada}, especially for the regional languages, dialects, and code-mixed speech common in linguistically diverse regions such as South Asia~\cite{Garg2022Handling}, which are underrepresented in existing corpora and pre-trained models.

The dominant strategy transcribes speech via ASR before applying text-based classifiers~\cite{sharon2022mada}, but this pipeline is ill-suited to low-resource audio abuse detection: ASR error rates are high for informal, code-mixed, profanity-rich speech~\cite{Chhabra2023A}; paralinguistic cues such as pitch and prosody that often signal abuse are lost in transcription~\cite{szekely2023prosody}; and cascaded errors increase both latency and failure risk~\cite{Tundik2019Assessing}. Audio-native models that avoid transcription altogether are therefore needed, but for multilingual, low-resource settings, it remains unclear how much labelled supervision such models actually require to be effective.

We address this gap using Contrastive Language--Audio Pre-training (CLAP)~\cite{elizalde2023clap}, which aligns audio and text in a shared embedding space and has shown strong zero-shot transfer in vision-language settings~\cite{Wang2025CLIP6D,Qian2024Online}. CLAP-based abuse detection is evaluated across ten Indic languages in the ADIMA benchmark under three supervision regimes: zero-shot prompt similarity (no labelled audio at all), classification on frozen CLAP embeddings, and few-shot supervised contrastive adaptation of the embedding space (at most 50 labelled examples per class), across monolingual, cross-lingual, and leave-one-language-out settings.

Any labelled supervision closes most of the gap to full supervision: classification on frozen, unadapted CLAP embeddings reaches 76.6 mean cross-lingual macro-F1, within 1--3 points of the fully supervised ADIMA benchmark~\citep{gupta2022adima}, and far above the 53.6--65.6 achieved by zero-shot prompting alone. Few-shot contrastive adaptation, by contrast, offers only a small and language-dependent additional benefit once support-set sampling variance is accounted for. The study is guided by the following research questions:

\begin{itemize}
    \item \textbf{RQ1:} How effective are CLAP-based representations for abusive speech detection directly from audio across multiple Indic languages?
    \item \textbf{RQ2:} To what extent does few-shot contrastive adaptation improve performance over using frozen pre-trained CLAP embeddings in low-resource language settings?
    \item \textbf{RQ3:} How well do CLAP embeddings transfer across languages, and does excluding the target language from training (leave-one-language-out) substantially degrade performance?
    \item \textbf{RQ4:} Does lightweight projection-only adaptation offer a more favourable trade-off between adaptation cost and downstream performance than deeper encoder fine-tuning in few-shot regimes?
\end{itemize}

ADIMA's fully supervised results~\citep{gupta2022adima} serve throughout as an upper-bound reference rather than a directly comparable few-shot baseline; the aim is to characterise how contrastive audio-text representations behave under varying degrees of supervision and cross-lingual transfer, not to maximise absolute performance on ADIMA.

\section{Literature Review}

Research on hate, abuse, and toxicity detection has been dominated by text-based approaches, but the growth of voice-based social media has exposed the limitations of text-only moderation~\cite{mnassri2022bert,jafari2023fine,mozafari2020bert}. Recent work has therefore begun to focus on direct speech-based detection, motivated by ASR brittleness in noisy, multilingual, and code-mixed conditions, as well as by latency and privacy concerns.

\subsection{Datasets for Audio Hate and Abuse Detection}

Labelled resources for audio hate speech detection remain limited in both size and linguistic coverage. Among the most important benchmarks is \textbf{ADIMA}~\cite{gupta2022adima}, which contains 11{,}775 audio clips spanning ten Indic languages and supports monolingual, multilingual, and cross-lingual evaluation, and highlights the limitations of ASR-based pipelines in profanity-rich and code-mixed speech.

Several other datasets extend this space: \textbf{DeToxy}~\cite{ghosh2022detoxy} and \textbf{MuTox}~\cite{costa2024mutox} increase multilingual coverage and show that direct audio-based systems can match or exceed ASR cascades; smaller language-specific datasets for Malayalam and Tamil remain useful but limited in scale~\cite{natarajan2025multimodal}; and \textbf{SynHate}~\cite{ranjan2025synhate} and \textbf{ToxicTone}~\cite{luo25_interspeech} extend evaluation to synthetic/deepfake audio and other languages. Taken together, these datasets reveal substantial heterogeneity in annotation schemes, language balance, and task formulation.

\subsection{Methodological Approaches}

Early audio abuse detection systems relied on handcrafted acoustic descriptors combined with traditional classifiers~\cite{rawat2023comprehensive,banuroopa2021mfcc}, but these transfer poorly across languages and domains. More recent work increasingly uses pre-trained speech encoders such as Wav2Vec~2.0, XLS-R, and Whisper, either as fixed feature extractors or fine-tuned backbones; on datasets such as ADIMA and MuTox, these audio-native representations can match or exceed ASR-based cascades, which remain vulnerable to transcription errors and added latency~\cite{sharon2022mada}.

Multimodal approaches combining acoustic and textual signals report gains over unimodal baselines~\cite{sharon2022mada,liu24h_interspeech}, and prosodic- or acoustic-only cues alone can remain discriminative without transcripts~\cite{spiesberger23_interspeech}. Shared tasks such as DravidianLangTech@NAACL 2025~\cite{chakravarthi2025proceedings} underscore the need for multimodal, multilingual moderation systems, while parallel work addresses robustness and interpretability under synthetic or adversarial perturbations~\cite{an2024explainable,ranjan2025synhate,ranjan25b_interspeech}.

\subsection{Few-Shot Learning and Low-Resource Challenges}

Low-resource audio abuse detection remains difficult because labelled data is scarce, annotation is expensive, and the notion of abuse is often culturally sensitive, with performance further affected by variability in speakers, accents, recording conditions, and code-mixing~\cite{gupta2022adima, sharon2022mada}. Few-shot approaches remain relatively underexplored here, though recent work on cross-lingual meta-learning over ADIMA shows promising adaptation behaviour alongside clear degradation in very low-shot regimes~\cite{sankaran2025crosslingual}.

In multilingual settings, transfer is complicated by dialectal variation, language family differences, script differences, and culturally specific notions of offensiveness; synthetic data generation and cross-lingual augmentation offer one route to scale but raise questions of fairness and cultural validity~\cite{Narula2024A,Nandi2024Combining,Roy2022Hate}. This appears to be the first work to apply few-shot supervised contrastive learning to audio-based hate/abuse speech detection, and among the first to apply CLAP-based audio-text representations to this task in Indic languages specifically.

\begin{figure*}[t]
    \centering
    \includegraphics[
        width=0.8\textwidth,
        trim=0 1.2cm 0 1.9cm,
        clip
    ]{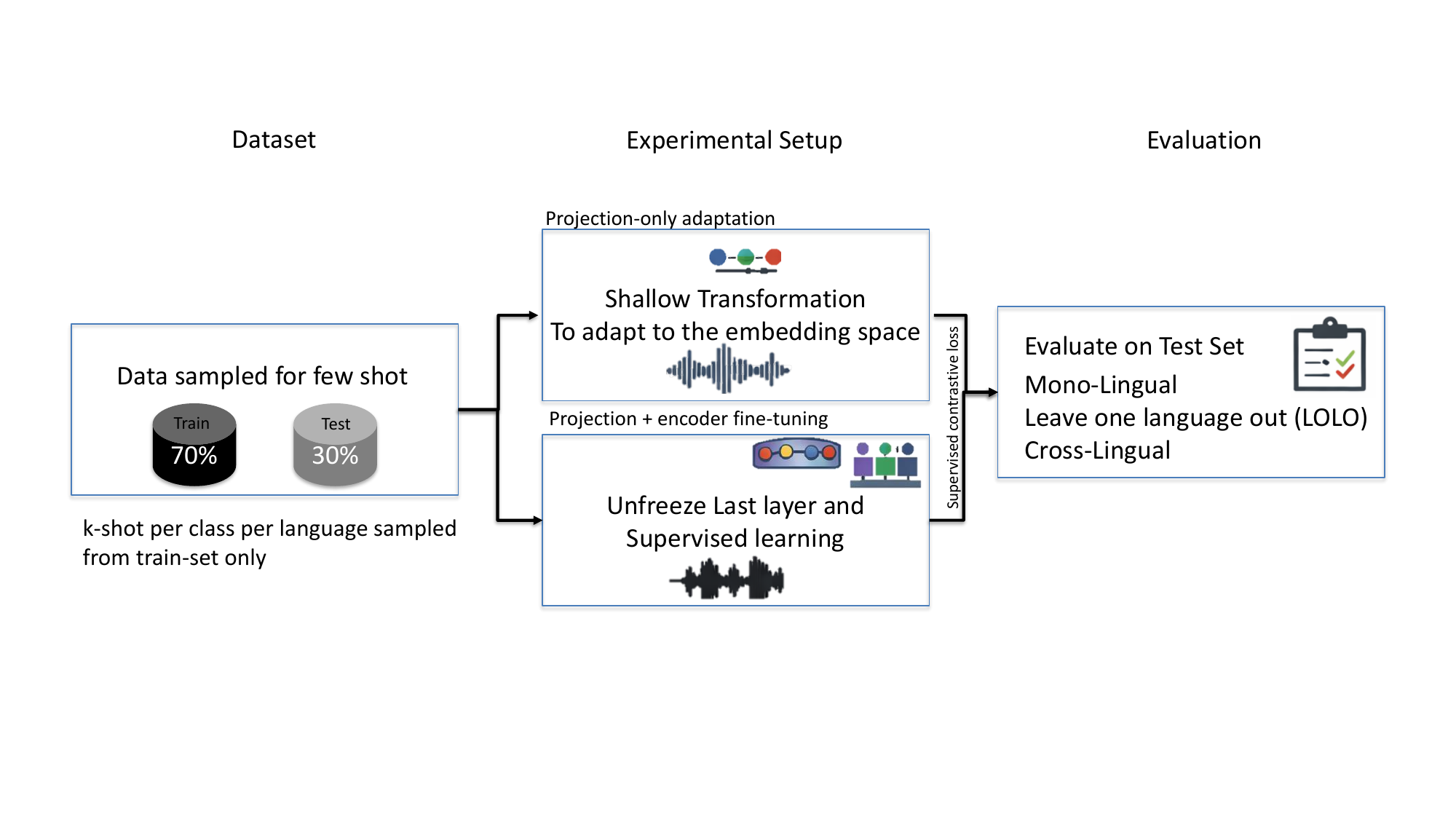}
    \caption{Overview of the proposed CLAP-based pipeline: a $k$-shot subset adapts CLAP via supervised contrastive learning (projection-only, or projection+fine-tuning), and the resulting embeddings are evaluated under cross-lingual and leave-one-language-out settings.}
    \label{fig:overall-method}
\end{figure*}

\section{Methodology}
\label{sec:method}

\subsection{Problem Formulation}

This work studies abusive speech detection from audio under \emph{zero-shot} and \emph{few-shot} conditions. Given an audio segment $x \in \mathcal{X}$ and a binary label $y \in \{0,1\}$ indicating non-abusive or abusive speech, the goal is to learn a classifier $f:\mathcal{X}\rightarrow\{0,1\}$ that generalises across languages $\ell \in \mathcal{L}$. Zero-shot inference uses no task-specific labelled audio, whereas few-shot learning assumes access to $k$ labelled examples per class per language drawn from the training split. The problem is cast as one of aligning audio signals with label semantics in a shared contrastive embedding space.

It is worth stressing that \emph{zero-shot} in this strict sense refers only to the prompt-similarity evaluation of Section~\ref{sec:prompt-sensitivity}, where no labelled audio is used at any stage. The ``$k{=}0$'' condition reported throughout the main results tables is different: it denotes a downstream classifier trained on the full labelled training split using \emph{frozen, unadapted} CLAP embeddings, where $k{=}0$ refers only to the number of examples used for contrastive adaptation of the embedding space, not to the supervision available to the classifier. This condition should not be conflated with the stricter zero-shot setting above.

\subsection{CLIP and CLAP}

CLIP~\cite{radford2021clip} aligns image and text embeddings using a contrastive objective, while CLAP~\cite{elizalde2023clap} replaces the image encoder with an audio encoder and learns joint audio-text representations. For a batch of $N$ aligned audio-text pairs $(a_i,t_i)$, with normalised embeddings $z^a_i$ and $z^t_i$, CLAP typically optimises a symmetric InfoNCE loss~\cite{oord2019infoceloss}:
\begin{equation}
\label{eq:infonce}
\mathcal{L}=\frac{1}{2}\left(\mathcal{L}_{a\rightarrow t}+\mathcal{L}_{t\rightarrow a}\right),
\end{equation}
\begin{equation}
\label{eq:infonce_a2t}
\mathcal{L}_{a\rightarrow t}=-\frac{1}{N}\sum_{i=1}^{N}\log
\frac{\exp(\mathrm{sim}(z^a_i,z^t_i)/\tau)}
{\sum_{j=1}^{N}\exp(\mathrm{sim}(z^a_i,z^t_j)/\tau)},
\end{equation}
with $\mathcal{L}_{t\rightarrow a}$ defined analogously. Here, $\mathrm{sim}(\cdot,\cdot)$ is cosine similarity and $\tau$ is a temperature parameter. Because label descriptions can be encoded as text, this framework naturally supports zero-shot classification.

\subsection{Audio Representation with CLAP}

A pre-trained CLAP audio encoder is used to extract $d$-dimensional audio embeddings, with $d=512$ in this implementation. Specifically, \texttt{clap-htsat-unfused}~\cite{elizalde2023clap} is used, which combines an HTSAT audio encoder with a RoBERTa text encoder, pre-trained exclusively on English audio--text pairs (LAION-Audio-630K and AudioSet). None of the ten target Indic languages appears in pre-training, so any observed cross-lingual transfer reflects generalisation of acoustic and paralinguistic structure rather than language overlap. Audio is resampled and padded or truncated to a fixed duration before encoding:
\begin{equation}
\label{eq:audio_embed}
z^a=\frac{f_{\mathrm{audio}}(x)}{\lVert f_{\mathrm{audio}}(x)\rVert_2}.
\end{equation}
These embeddings are used in three ways: for zero-shot prompt similarity, for few-shot supervised contrastive adaptation, and as features for downstream supervised classifiers.

\subsection{Zero-shot Prompting}

For zero-shot classification, two prompts $p_0$ and $p_1$ are defined, representing the non-abusive and abusive classes, respectively, for example:
\emph{``This audio contains hate speech''} and \emph{``This audio does not contain hate speech''}. Normalised text embeddings are then computed:
\begin{equation}
\label{eq:text_embed}
z^t_c=\frac{f_{\mathrm{text}}(p_c)}{\lVert f_{\mathrm{text}}(p_c)\rVert_2},
\end{equation}
and predict the label using prompt similarity:
\begin{equation}
\label{eq:zs_pred}
\hat{y}=\arg\max_{c\in\{0,1\}}\ \mathrm{sim}(z^a,z^t_c).
\end{equation}

\subsection{Few-Shot Contrastive Adaptation}

To adapt CLAP to abusive speech detection in low-resource conditions, support sets are constructed with $k$ examples per class per language from the training data. These few-shot examples are used to refine the embedding space through supervised contrastive learning, under two adaptation strategies. In \textbf{Projection-only adaptation}, the CLAP encoder is frozen, and only a lightweight projection head is trained. In \textbf{Projection+fine-tuning}, the projection head is trained jointly with the final audio encoder blocks~\cite{Liang2024AAT,Kim2022Integrated}.

Given projected embeddings $u_i = g(z^a_i)$, the supervised contrastive objective is optimised, based on ~\cite{khosla2020supervised}:
\begin{equation}
\label{eq:supcon}
\begin{aligned}
\mathcal{L}_{\mathrm{sup}}
&= \sum_{i \in I} \frac{-1}{|P(i)|}
\sum_{p \in P(i)}
\log \\
&\quad \frac{
\exp\left(\mathrm{sim}(u_i,u_p)/\tau\right)
}{
\sum_{a \in A(i)}
\exp\left(\mathrm{sim}(u_i,u_a)/\tau\right)
},
\end{aligned}
\end{equation}
where $P(i)$ denotes the set of positive examples sharing the same label as $i$, and $A(i)$ denotes all other samples in the batch. This loss encourages embeddings from the same class to form tighter clusters while increasing separation between classes. After adaptation, the resulting embeddings are used for downstream classification.

\subsection{Downstream Classifiers}

Two downstream classifiers are evaluated on frozen and adapted embeddings: a Support Vector Machine (SVM) and a scikit-learn ANN-MLP. These models make it possible to assess whether abuse-related information is linearly or non-linearly separable in the CLAP embedding space, and whether few-shot contrastive adaptation improves downstream decision boundaries.

\paragraph{Classifier selection}

For each language and evaluation setting, both the SVM and the ANN-MLP are trained on the adapted or frozen CLAP embeddings, and the higher macro-F1 result is selected, reported consistently across all tables without varying between shot sizes for a given language. This strategy is adopted because no single downstream classifier is uniformly strongest across all languages and transfer settings, a pattern consistent with prior work on embedding-based classification in low-resource multilingual settings~\cite{Unanue2023T3L,Raza2025A}.

\section{Experiments}
\label{sec:experiments}

\subsection{Data, Splits, and Preprocessing}

Each instance consists of a waveform $x$, a binary abuse label $y \in \{0,1\}$, and a language tag $\ell \in \mathcal{L}$. Audio is resampled and padded or truncated to a fixed duration before CLAP feature extraction. The train/test partition provided with the ADIMA dataset~\cite{gupta2022adima} is followed, where available, and preserved throughout all experiments. For evaluations that condition on language, training and testing are performed within the corresponding language partitions. Few-shot support sets are always sampled only from the training split, ensuring that no test information leaks into adaptation.

{\renewcommand{\arraystretch}{1.2}
\begin{table*}[t]
\centering
\small
\setlength{\tabcolsep}{6pt}
\begin{tabular}{l|c|c|c}
\hline
\textbf{Language} & \textbf{ADIMA (0-shot)} & \textbf{Projection-only} & \textbf{Projection+FT} \\
\hline
Bengali    & \textbf{79.10} & 76.3 (0-shot)  & 76.3 (0-shot) \\
Bhojpuri   & --             & 70.5 (5-shot)  & 70.5 (50-shot) \\
Gujarati   & --             & \textbf{74.6} (0-shot) & \textbf{74.6} (0-shot) \\
Haryanvi   & --             & \textbf{77.6} (10-shot) & 77.4 (0-shot) \\
Hindi      & \textbf{80.70} & 77.8 (0-shot)  & 77.8 (0-shot) \\
Kannada    & \textbf{78.40} & 76.8 (0-shot)  & 76.8 (0-shot) \\
Malayalam  & --             & \textbf{77.9} (5-shot) & 77.6 (5-shot) \\
Odia       & --             & \textbf{81.5} (0-shot) & \textbf{81.5} (0-shot) \\
Punjabi    & \textbf{83.40} & 82.3 (0-shot)  & 82.3 (0-shot) \\
Tamil      & \textbf{75.20} & 73.0 (0-shot)  & 73.0 (0-shot) \\
\hline
\end{tabular}
\caption{Cross-lingual macro-F1 (\%) on ADIMA. Projection-only and projection+fine-tuning report mean macro-F1 over 5 support-set draws at the best-performing shot size (single run at $k=0$), shown in parentheses; ADIMA results from~\citet{gupta2022adima} shown where available. Best score per row in bold.}
\label{tab:crosslingual_mf1_main}
\end{table*}
} % end \renewcommand{\arraystretch}

\subsection{Experimental Setups}

The proposed framework is evaluated along two axes: \emph{language generalisation setting} and \emph{representation setting}, frozen CLAP embeddings (Eq.~\ref{eq:audio_embed}), few-shot supervised contrastive adaptation on $k$-shot support sets, or zero-shot prompting (Eq.~\ref{eq:zs_pred}), using an SVM and a scikit-learn ANN-MLP as downstream classifiers on the resulting embeddings. Zero-shot prompting is treated as an auxiliary evaluation setting; the main results concern downstream classification on frozen versus few-shot-adapted embeddings. The benefit of adaptation over frozen 0-shot embeddings is language-dependent rather than uniform: for the majority of languages (e.g.\ Bengali, Gujarati, Hindi, Kannada, Odia, Punjabi, Tamil) the frozen embedding space is already near-optimal, whereas Bhojpuri shows a clear, variance-robust gain from adaptation (Section~\ref{sec:variance}).

\noindent\textbf{Language generalisation settings:}
\begin{itemize}
    \item \textbf{Monolingual:} for each language $\ell$, train on $\ell_{\text{train}}$ and test on $\ell_{\text{test}}$.
    \item \textbf{Leave-one-language-out (LOLO):} for each target language $\ell^\star$, train on $\bigcup_{\ell\neq\ell^\star}\ell_{\text{train}}$ and test on $\ell^\star_{\text{test}}$.
    \item \textbf{Cross-lingual / joint multilingual:} train on $\bigcup_{\ell\in\mathcal{L}}\ell_{\text{train}}$ and report results per target language on $\ell_{\text{test}}$.
\end{itemize}

\noindent\textbf{Representation settings:}
\begin{itemize}
    \item \textbf{Frozen embeddings:} train downstream classifiers directly on pre-trained CLAP audio embeddings.
    \item \textbf{Few-shot adapted embeddings:} adapt CLAP representations using supervised contrastive learning on a small support set, then evaluate downstream classifiers on the adapted embeddings.
    \item \textbf{Zero-shot prompting:} classify audio using similarity to abusive and non-abusive text prompts.
\end{itemize}

\subsection{Implementation and Metrics}
% ~\footnote{Code attached as supplementary material}
CLAP feature extraction and adaptation are implemented in PyTorch and HuggingFace using the Supervised Contrastive Loss~\cite{khosla2020supervised}. The downstream SVM and ANN-MLP classifiers are implemented with scikit-learn. Random seeds are fixed for reproducibility, including few-shot sampling. Accuracy and Macro-F1 are reported, with language-wise results across languages. Since the original ADIMA paper~\cite{gupta2022adima} reported Macro-F1, it is used as the main comparison metric in this paper. 0-, 1-, 5-, 10-, 25-, and 50-shot settings are evaluated for supervised contrastive adaptation. Full hyper-parameter settings are provided in Appendix~\ref{app:hyperparams}.

\section{Results}

\subsection{Cross-lingual Results}

Table~\ref{tab:crosslingual_mf1_main} reports cross-lingual macro-F1 for the \textbf{projection-only} and \textbf{projection+fine-tuning} settings, alongside ADIMA's fully supervised results~\citep{gupta2022adima} where available (mean over 5 support-set draws at the best-performing shot size; single deterministic run at 0-shot). ADIMA reflects full per-language supervision, whereas the proposed framework adapts the embedding space with at most 50 labelled examples per class; the downstream classifier is nonetheless always trained on the full labelled training split (Section~\ref{sec:method}). The ADIMA figures are therefore an upper-bound reference rather than a directly comparable few-shot baseline, and the degree to which this approach closes the gap to full supervision is the relevant measure of performance.

Once macro-F1 is averaged over 5 seeds rather than taken from a single run, ADIMA is the strongest system for every language with a comparison available: Bengali (76.3 vs.\ 79.10), Hindi (77.8 vs.\ 80.70), Kannada (76.8 vs.\ 78.40), Punjabi (82.3 vs.\ 83.40), and Tamil (73.0 vs.\ 75.20). An earlier single-run analysis had reported Punjabi's 5-shot result (83.65) as marginally surpassing ADIMA; under the 5-seed mean, this becomes 82.2$\pm$0.9, and the best mean over any shot size (82.3, at 0-shot) remains 1.1 points below ADIMA. That earlier claim is withdrawn: the correct reading is that projection-only adaptation, and, for most languages, even frozen 0-shot embeddings alone, comes within roughly 1--3 macro-F1 points of full supervision, a practically meaningful result where full annotation is infeasible.

\paragraph{Where does adaptation help?} Selecting the best shot size by multi-seed mean rather than a single draw (Table~\ref{tab:variance}) changes this picture substantially. Table~\ref{tab:shot-optimal} (Appendix~\ref{app:extra_tabs}) shows the frozen 0-shot embedding is best for seven of the ten languages: Bengali, Gujarati, Hindi, Kannada, Odia, Punjabi, and Tamil. Only Bhojpuri shows a clear, variance-robust gain (68.9 at 0-shot vs.\ 70.5$\pm$0.7 at 5-shot, over two standard deviations above 0-shot); Malayalam's gain is smaller and borderline (76.9 vs.\ 77.9$\pm$1.0, roughly one standard deviation), and Haryanvi's nominal best (77.6$\pm$0.8 at 10-shot) lies within one standard deviation of its 0-shot score and is not reliably distinguishable from it. For the large majority of languages, lightweight supervised contrastive adaptation does not improve on frozen CLAP embeddings once sampling variance is taken into account.

\paragraph{Non-monotonic shot-size behaviour.} Performance does not increase monotonically with shot size. At very small $k$ (especially $k{=}1$), there are too few positive pairs to meaningfully reshape the embedding geometry; as $k$ grows, the projection head can instead overfit to the support set rather than learning transferable clusters. For languages whose frozen embedding space is already well structured (Table~\ref{tab:shot-optimal}), adaptation therefore risks degradation rather than gain, the dominant pattern in these results, with Bhojpuri the clearest exception. This explains why the optimal shot size varies across languages and why the mean curves in Figures~\ref{fig:lolo_mean_a} and~\ref{fig:lolo_mean_b} do not increase monotonically.

\subsection{Leave-One-Language-Out Analysis}
\label{sec:lolo_discussion}

To assess how much performance depends on target-language supervision, the standard cross-lingual setting (target language included in multilingual training data) is compared against a leave-one-language-out (LOLO) setup (target language fully excluded, used only for testing). If performance remains stable under LOLO, the model is likely capturing language-agnostic cues of abusive speech rather than relying mainly on target-language-specific supervision.

\begin{figure*}[t]
    \centering
    \begin{subfigure}[t]{0.48\textwidth}
        \centering
        \includegraphics[width=\textwidth]{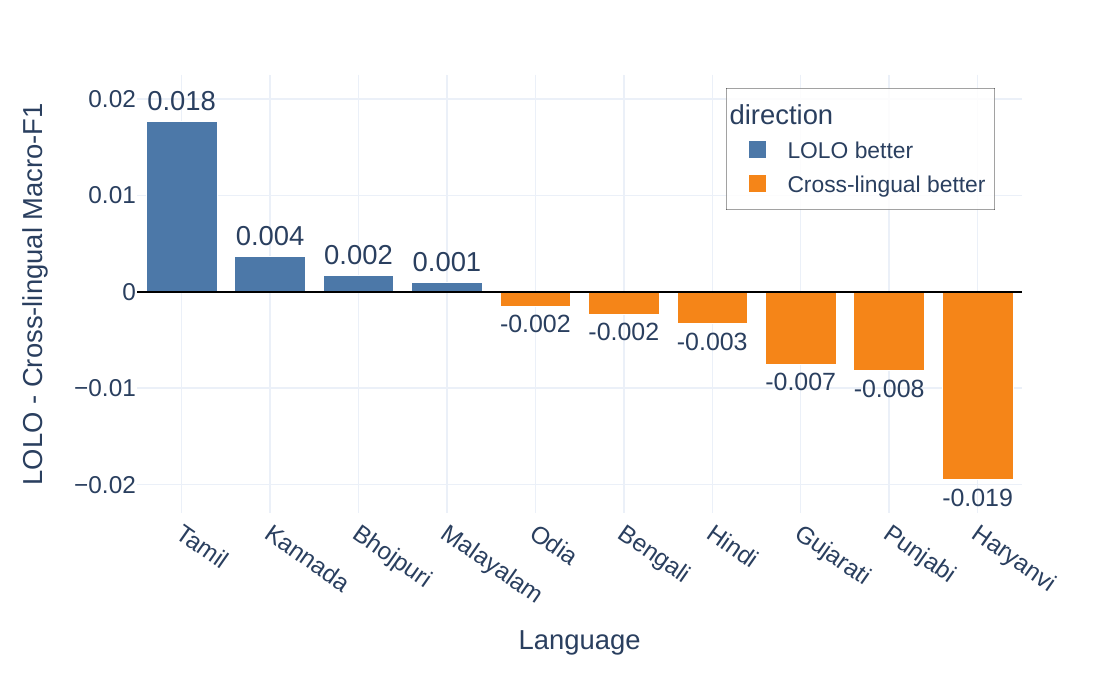}
        \caption{Proj.-only: LOLO$-$cross-lingual macro-F1 difference.}
        \label{fig:lolo_delta_a}
    \end{subfigure}
    \hfill
    \begin{subfigure}[t]{0.48\textwidth}
        \centering
        \includegraphics[width=\textwidth]{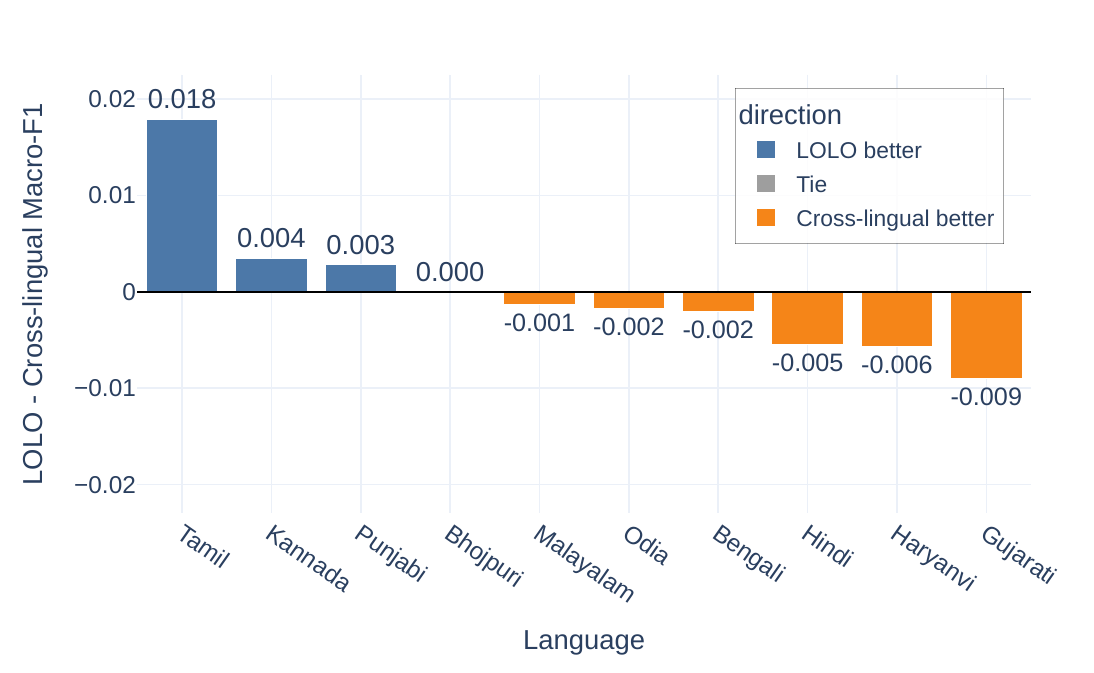}
        \caption{Proj.+FT: LOLO$-$cross-lingual macro-F1 difference.}
        \label{fig:lolo_delta_b}
    \end{subfigure}

    \vspace{0.4em}
    \begin{subfigure}[t]{0.48\textwidth}
        \centering
        \includegraphics[width=\textwidth]{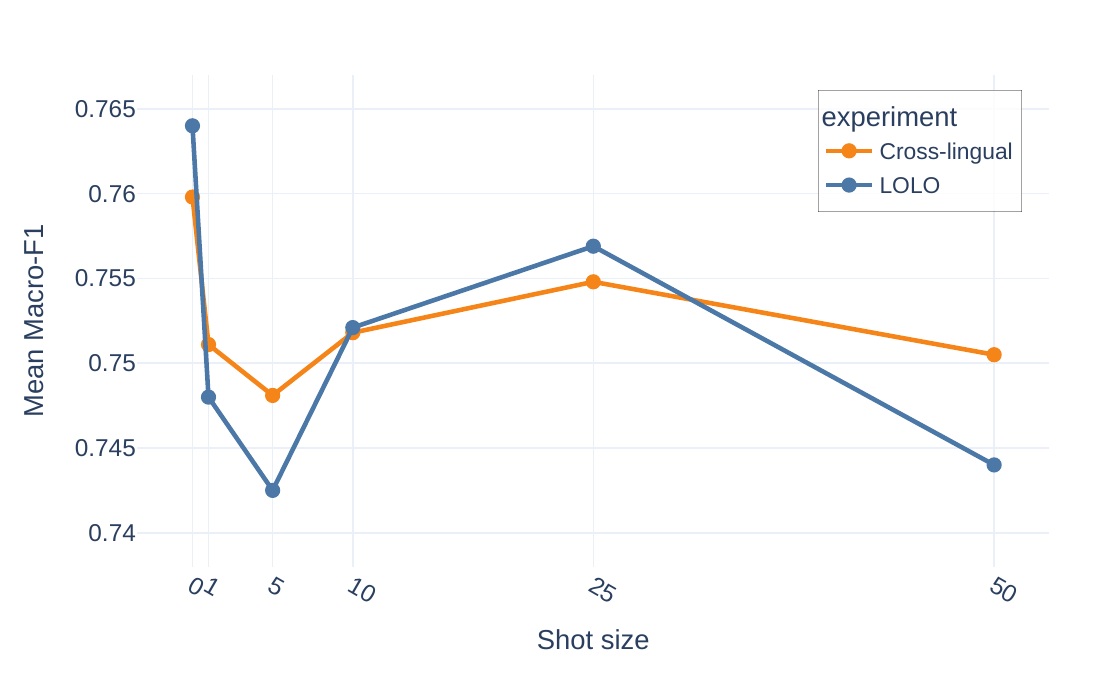}
        \caption{Proj.-only: mean macro-F1 by shot size.}
        \label{fig:lolo_mean_a}
    \end{subfigure}
    \hfill
    \begin{subfigure}[t]{0.48\textwidth}
        \centering
        \includegraphics[width=\textwidth]{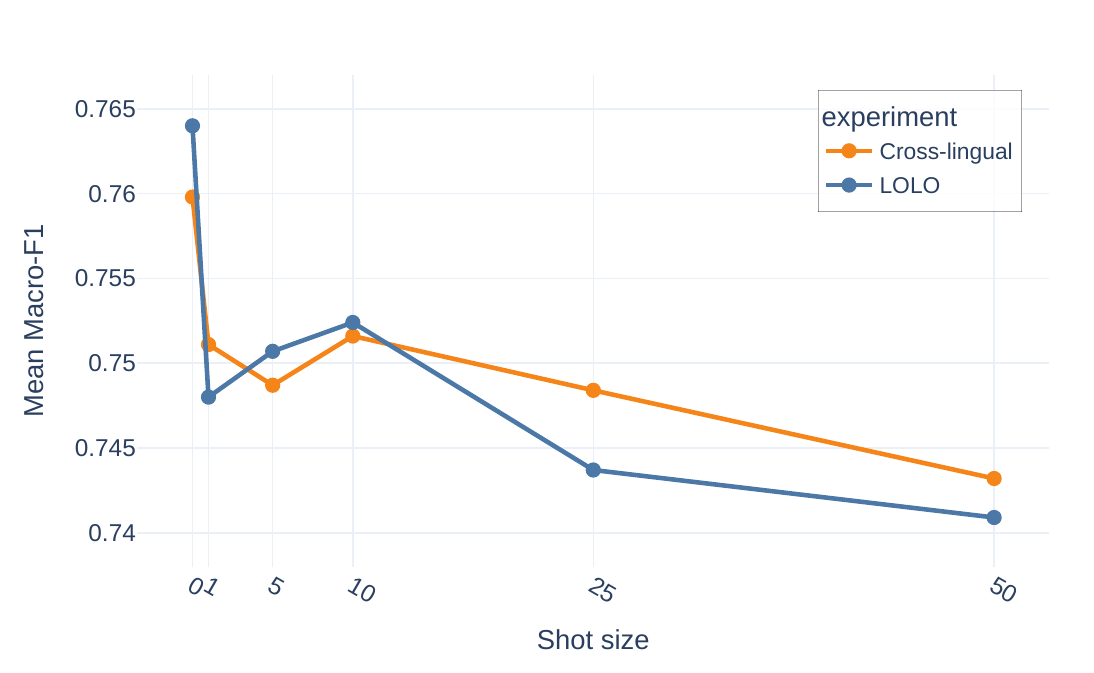}
        \caption{Proj.+FT: mean macro-F1 by shot size.}
        \label{fig:lolo_mean_b}
    \end{subfigure}
    \caption{LOLO vs.\ cross-lingual evaluation. Top: per-language macro-F1 difference (positive = LOLO stronger). Bottom: mean macro-F1 by shot size.}
    \label{fig:lolo_combined}
\end{figure*}

Figure~\ref{fig:lolo_combined} shows that, in both adaptation settings, \textbf{LOLO remains competitive with cross-lingual evaluation but does not outperform it on average}: cross-lingual training retains a small, persistent advantage, indicating that direct exposure to the target language still provides useful information. Language-wise, LOLO outperforms cross-lingual for a subset of languages, most clearly \textit{Tamil}, with smaller gains for \textit{Kannada} and \textit{Bhojpuri}, and for \textit{Punjabi} under projection+fine-tuning, suggesting multilingual transfer can generalise well even without target-language training data. Cross-lingual instead remains clearly stronger for \textit{Haryanvi} and \textit{Gujarati}, indicating these languages benefit more from direct target-language exposure. The LOLO--cross-lingual gap is also small and unstable rather than monotonic across shot sizes, so target-language exclusion is neither uniformly harmful nor systematically beneficial.

LOLO is best interpreted not as a stronger alternative to cross-lingual training but as a \textbf{transfer analysis}: the small size of the LOLO--cross-lingual gap is itself evidence that the model learns partially language-agnostic representations of abusive speech, since excluding the target language from training costs little. The persistent, if small, average advantage of cross-lingual training indicates that this transfer is meaningful but incomplete, with the balance between transferable and language-specific information varying across languages.

\subsection{Variance across Support-Set Sampling}
\label{sec:variance}

A single support-set draw per shot size cannot distinguish a genuine trend from sampling noise. The $k$-shot support set is therefore resampled under $S{=}5$ seeds per shot size, with mean $\pm$ std macro-F1 reported (projection-only, cross-lingual; full per-language values in Table~\ref{tab:variance}, Appendix~\ref{app:extra_tabs}); the $k{=}0$ column has no associated variance, since frozen embeddings involve no support-set draw. Contrary to the initial expectation that variance would be largest at $k{=}1$ and contract as $k$ grows, the mean standard deviation across languages is broadly flat (1-shot: 0.8; 5-shot: 1.0; 10-shot: 1.0; 25-shot: 1.2; 50-shot: 0.8), with 25-shot marginally the widest; even at 50-shot, several languages (e.g.\ Bengali, Haryanvi) retain a standard deviation above 1 point of macro-F1.

\subsection{Isolating the CLAP Contribution}
\label{sec:encoder-ablation}

To test whether observed transfer stems from CLAP's contrastive audio--text grounding specifically, rather than from any sufficiently strong pre-trained encoder, the pipeline is held fixed (projection-only adaptation, identical classifiers and shot sizes) while the CLAP encoder is substituted with four alternatives: Whisper Large-v3~\citep{radford2022whisper} (weak ASR supervision, no contrastive audio--text objective); Wav2Vec~2.0-large and Wav2Vec~2.0-base (self-supervised masked prediction, no textual grounding, no exposure to any target language); and XLS-R-300m (multilingual self-supervised, pre-training corpus includes 7 of the 10 target languages). Table~\ref{tab:encoder-ablation} reports mean cross-lingual macro-F1 over 5 support-set draws for $k\geq1$ (single run at $k{=}0$).

Whisper matches or exceeds CLAP at every shot size, and its best few-shot mean (79.2 at 50-shot) is reliably above its own 0-shot score (77.1); CLAP's few-shot means do not improve on its 0-shot score. The three additional self-supervised encoders pattern like CLAP rather than Whisper: 0-shot is each one's best score, with no reliable few-shot gain (Wav2Vec2-large: 75.6 vs.\ 74.4; XLS-R: 76.2 vs.\ 75.3; Wav2Vec2-base: 77.2 vs.\ 76.0). XLS-R's partial target-language exposure gives it no clear edge over the English-only Wav2Vec2-base, so prior language exposure alone does not explain the differences between encoders either. Whisper , the only encoder here with any textual (ASR) supervision , remains the sole exception, pointing to ASR-style supervision or scale, rather than a contrastive audio--text objective, as the likelier source of its distinct behaviour. Three self-supervised encoders with no textual grounding at all thus reproduce CLAP's own pattern of static few-shot performance: this is evidently not a weakness specific to CLAP's pre-training, and whether few-shot contrastive adaptation reliably improves on any frozen encoder tested here remains open (Section~\ref{sec:variance}).

{\renewcommand{\arraystretch}{1.2}
\begin{table}[t]
\centering
\small
\resizebox{\linewidth}{!}{%
\begin{tabular}{@{}lccc@{}}
\toprule
Encoder & 0-shot & best few-shot & avg \\
\midrule
CLAP (\texttt{clap-htsat-unfused}) & 76.6 & 75.9 (1-shot) & 75.9 \\
Whisper Large-v3                   & 77.1 & 79.2 (50-shot) & 78.4 \\
Wav2Vec2-large           & 75.6 & 74.4 (1-shot)  & 74.3 \\
XLS-R-300m$^\dagger$     & 76.2 & 75.3 (50-shot) & 75.1 \\
Wav2Vec2-base            & 77.2 & 76.0 (1-shot)  & 76.0 \\
\bottomrule
\end{tabular}%
}
\caption{Encoder ablation, projection-only (cross-lingual mean macro-F1 over 5 draws for $k\geq1$; single run at $k=0$). ``avg'' is the mean over all six shot sizes. $\dagger$: XLS-R's pre-training includes 7 of the 10 target languages, so cannot cleanly isolate the SSL objective from prior exposure.}
\label{tab:encoder-ablation}
\end{table}
} % end \renewcommand{\arraystretch}

\subsection{Zero-Shot Prompt Sensitivity}
\label{sec:prompt-sensitivity}

Because the text encoder operates entirely in English semantic space while the audio is Indic-language speech, zero-shot performance depends on how language-agnostic the paralinguistic cues of abuse are, and on prompt wording. Four prompt formulations are evaluated (Table~\ref{tab:prompt-sensitivity}): \emph{P1\_basic} (``This audio contains abusive speech.'' / ``\ldots normal speech.''), \emph{P2\_hate} (``This is an example of verbal abuse or hate speech.'' / ``\ldots normal, non-offensive conversation.''), \emph{P3\_offensive} (``The speaker is using offensive or hateful language.'' / ``\ldots polite and neutral language.''), and \emph{P4\_aggressive} (``Audio of someone being abusive, aggressive, or hateful.'' / ``\ldots calm, respectful conversation.'').

Prompt wording substantially affects zero-shot macro-F1: the elaborated hate/offence prompts (P2, P3) outperform the literal P1\_basic wording by roughly 12 points on average, and stability is highly language-dependent (Bhojpuri/Bengali most stable at 7.2--8.3 points range; Malayalam/Punjabi/Tamil least stable at 16.9--19.2). This also clarifies where the framework's practical value comes from: true zero-shot prompting tops out at 65.6 mean macro-F1 (P2\_hate) and is as low as 53.6 (P1\_basic), whereas a classifier on frozen, unadapted CLAP embeddings alone already reaches 76.6 (Table~\ref{tab:encoder-ablation}, $k{=}0$) , a gap of 11--23 points that holds for every encoder in this ablation (75.6--77.2, Table~\ref{tab:encoder-ablation}). The decisive step is therefore not few-shot contrastive adaptation, but the move from no labelled supervision to any labelled supervision at all; whether that supervision also reshapes the embedding space via few-shot adaptation is, as Section~\ref{sec:variance} shows, a secondary and largely unreliable source of further gains. Full per-prompt mean/min/max values are given in Table~\ref{tab:prompt-sensitivity} (Appendix~\ref{app:extra_tabs}).

\section{Conclusion}

This paper investigated whether CLAP-based audio-text representations support multilingual abusive speech detection directly from audio, in low-resource and cross-lingual settings. CLAP provides a strong representation space across the ADIMA benchmark, with projection-only adaptation reaching a cross-lingual average macro-F1 of \textbf{76.83} (mean over 5 draws). ADIMA remains the stronger system for every language with a full-supervision comparison, including Punjabi (82.3 vs.\ 83.40); an earlier single-run claim that adaptation surpassed ADIMA for Punjabi at 5-shot did not hold up under multi-seed re-analysis (Section~\ref{sec:variance}).

Compared to true zero-shot prompting (53.6--65.6 mean macro-F1), frozen CLAP embeddings alone reach 76.6: any labelled supervision matters far more than whether it also reshapes the embedding space. Few-shot adaptation itself is largely unreliable once variance is accounted for , seven of ten languages are best at \textbf{0-shot} (Table~\ref{tab:shot-optimal}), with \textbf{Bhojpuri} the clear exception.

LOLO remained close to cross-lingual performance (76.81/76.68 vs.\ 76.83/76.78 average macro-F1), confirming meaningful cross-lingual structure. Contrastive audio-text models are a promising basis for multilingual audio abuse detection despite the limited reliability of few-shot adaptation; future work should explore prompt design, parameter-efficient adaptation, and multimodal fusion.

\section*{Limitations}

These findings should be interpreted in light of several limitations. First, the experiments are conducted on a single multilingual benchmark, ADIMA, so the results may not generalise directly to other datasets, recording conditions, or non-Indic languages. Second, although CLAP is motivated partly by its zero-shot capability, the strongest results in this paper come from few-shot adaptation and downstream classification; zero-shot prompting is therefore only an auxiliary evaluation setting. Third, the few-shot results may be sensitive to support-set composition and optimisation choices, particularly in very low-shot regimes~(1-shot and 5-shot). Random seeds are fixed for reproducibility, and results are reported across six shot sizes~($k \in \{0, 1, 5, 10, 25, 50\}$) rather than relying on a single condition, which partially mitigates this concern by revealing whether performance trends are consistent across data regimes. Macro-F1 is now reported as a mean with standard deviation over 5 randomly sampled support sets per shot size (Table~\ref{tab:variance}), which quantifies support-set sampling variance directly. This variance is substantial and does not shrink monotonically with shot size as initially anticipated; at every shot size, several languages have a standard deviation exceeding 1 point of macro-F1, and the frozen 0-shot embedding is the best or statistically indistinguishable from the best few-shot mean for most languages. Fourth, only two downstream classifiers are evaluated, SVM and ANN-MLP, and only two lightweight adaptation strategies are compared, so the conclusions do not extend to all possible fine-tuning or parameter-efficient methods. Fifth, the projection head architecture (a 2-layer MLP with a 256-dimensional output) is fixed throughout and follows established practice from supervised contrastive learning~\citep{khosla2020supervised} rather than being ablated over depth, width, or activation choice; a systematic ablation of head design is left to future work. Finally, this study does not include a dedicated fairness analysis across speakers, dialects, or culturally specific abusive expressions.

\section*{Ethics Statement}
This study does not involve any personal or public data pointing to an individual or a group of individuals and thus does not break any ethical guidelines.

\bibliography{mybib}

@inproceedings{gupta2022adima,
  author    = {Vikram Gupta and Rini Sharon and Ramit Sawhney and Debdoot Mukherjee},
  title     = {{ADIMA}: Abuse Detection in Multilingual Audio},
  booktitle = {Proceedings of ICASSP 2022},
  year      = {2022},
  pages     = {6172--6176},
  note      = {arXiv:2202.07991}
}

@inproceedings{ghosh2022detoxy,
  author    = {Sreyan Ghosh and Samden Lepcha and Sakshi S. and Rajiv Ratn Shah and Srinivasan Umesh},
  title     = {{DeToxy}: A Large-Scale Dataset for Toxicity Classification in Spoken Utterances},
  booktitle = {Proceedings of Interspeech 2022},
  year      = {2022},
  pages     = {5185--5189}
}

@inproceedings{sharon2022mada,
  author    = {Rini Sharon and Heet Shah and Debdoot Mukherjee and Vikram Gupta},
  title     = {Multilingual and Multimodal Abuse Detection},
  booktitle = {Proceedings of Interspeech 2022},
  year      = {2022}
}

@inproceedings{sankaran2025crosslingual,
  author    = {Aditya Narayan Sankaran and Reza Farahbakhsh and Noel Crespi},
  title     = {Towards Cross-Lingual Audio Abuse Detection in Low-Resource Settings with Few-Shot Learning},
  booktitle = {Proceedings of COLING 2025},
  year      = {2025}
}

@article{ranjan2025synhate,
  author  = {Rishabh Ranjan and Kishan Pipariya and Mayank Vatsa and Richa Singh},
  title   = {SynHate: Detecting Hate Speech in Synthetic Deepfake Audio},
  journal = {arXiv preprint arXiv:2506.06772},
  year    = {2025}
}

@article{an2024explainable,
  author  = {Jinmyeong An and Wonjun Lee and Yejin Jeon and Jungseul Ok and Yunsu Kim and Gary Geunbae Lee},
  title   = {An Investigation into Explainable Audio Hate Speech Detection},
  journal = {arXiv preprint arXiv:2408.06065},
  year    = {2024}
}

@inproceedings{costa2024mutox,
  title={Mutox: Universal multilingual audio-based toxicity dataset and zero-shot detector},
  author={Costa-juss{\`a}, Marta and Meglioli, Mariano and Andrews, Pierre and Dale, David and Hansanti, Prangthip and Kalbassi, Elahe and Mourachko, Alexandre and Ropers, Christophe and Wood, Carleigh},
  booktitle={Findings of the Association for Computational Linguistics: ACL 2024},
  pages={5725--5734},
  year={2024}
}

@inproceedings{natarajan2025multimodal,
  title={Multimodal Hate Speech Detection in Dravidian Languages Using Text and Audio Features},
  author={Natarajan, Radha and Swathika, R and others},
  booktitle={2025 11th International Conference on Communication and Signal Processing (ICCSP)},
  pages={148--152},
  year={2025},
  organization={IEEE}
}

@inproceedings{szekely2023prosody,
  title={Prosody-controllable Gender-ambiguous Speech Synthesis: A Tool for Investigating Implicit Bias in Speech Perception.},
  author={Sz{\'e}kely, {\'E}va and Gustafson, Joakim and Torre, Ilaria},
  booktitle={Interspeech},
  pages={1234--1238},
  year={2023}
}

@INPROCEEDINGS{mnassri2022bert,
  author={Mnassri, Khouloud and Rajapaksha, Praboda and Farahbakhsh, Reza and Crespi, Noel},
  booktitle={GLOBECOM 2022 - 2022 IEEE Global Communications Conference}, 
  title={BERT-based Ensemble Approaches for Hate Speech Detection}, 
  year={2022},
  volume={},
  number={},
  pages={4649-4654},
  doi={10.1109/GLOBECOM48099.2022.10001325}}

@ARTICLE{jafari2023fine,
  author={Jafari, Amir Reza and Li, Guanlin and Rajapaksha, Praboda and Farahbakhsh, Reza and Crespi, Noel},
  journal={IEEE Access}, 
  title={Fine-Grained Emotions Influence on Implicit Hate Speech Detection}, 
  year={2023},
  volume={11},
  number={},
  pages={105330-105343},
  doi={10.1109/ACCESS.2023.3318863}}

@InProceedings{mozafari2020bert,
author="Mozafari, Marzieh
and Farahbakhsh, Reza
and Crespi, No{\"e}l",
editor="Cherifi, Hocine
and Gaito, Sabrina
and Mendes, Jos{\'e} Fernendo
and Moro, Esteban
and Rocha, Luis Mateus",
title="A BERT-Based Transfer Learning Approach for Hate Speech Detection in Online Social Media",
booktitle="Complex Networks and Their Applications VIII",
year="2020",
publisher="Springer International Publishing",
address="Cham",
pages="928--940",
isbn="978-3-030-36687-2"
}

@article{Narula2024A,
title={A comprehensive review on detection of hate speech for multi-lingual data},
author={Rachna Narula and Poonam Chaudhary},
journal={Social Network Analysis and Mining},
year={2024},
volume={14},
doi={10.1007/s13278-024-01401-y}
}

@article{Nandi2024Combining,
title={Combining multiple pre-trained models for hate speech detection in Bengali, Marathi, and Hindi},
author={Arpan Nandi and Kamal Sarkar and Arjun Mallick and Arkadeep De},
journal={Multimedia Tools and Applications},
year={2024},
volume={83},
pages={77733 - 77757},
doi={10.1007/s11042-023-17934-x}
}

@article{Roy2022Hate,
title={Hate speech and offensive language detection in Dravidian languages using deep ensemble framework},
author={P. Roy and Snehaan Bhawal and C. N. Subalalitha},
journal={Comput. Speech Lang.},
year={2022},
volume={75},
pages={101386},
doi={10.1016/j.csl.2022.101386}
}

@inproceedings{chakravarthi2025proceedings,
  title={Proceedings of the Fifth Workshop on Speech, Vision, and Language Technologies for Dravidian Languages},
  author={Chakravarthi, Bharathi Raja and Priyadharshini, Ruba and Thavareesan, Sajeetha and Sherly, Elizabeth and Rajiakodi, Saranya and Palani, Balasubramanian and Subramanian, Malliga and Cn, Subalalitha and Chinnappa, Dhivya and others},
  booktitle={Proceedings of the Fifth Workshop on Speech, Vision, and Language Technologies for Dravidian Languages},
  year={2025}
}

@article{rawat2023comprehensive,
  title={A comprehensive study based on MFCC and spectrogram for audio classification},
  author={Rawat, Priyanshu and Bajaj, Madhvan and Vats, Satvik and Sharma, Vikrant},
  journal={Journal of Information and Optimization Sciences},
  volume={44},
  number={6},
  pages={1057--1074},
  year={2023}
}

@article{banuroopa2021mfcc,
  title={MFCC based hybrid fingerprinting method for audio classification through LSTM},
  author={Banuroopa, Kalyanaswamy and Shanmuga Priyaa, D},
  journal={International Journal of Nonlinear Analysis and Applications},
  volume={12},
  number={Special Issue},
  pages={2125--2136},
  year={2021},
  publisher={Semnan University}
}

@inproceedings{spiesberger23_interspeech,
  title     = {{Abusive Speech Detection in Indic Languages Using Acoustic Features}},
  author    = {Anika A. Spiesberger and Andreas Triantafyllopoulos and Iosif Tsangko and Bj{"o}rn W. Schuller},
  year      = {2023},
  booktitle = {{Interspeech 2023}},
  pages     = {2683--2687},
  doi       = {10.21437/Interspeech.2023-789}
}

@inproceedings{liu24h_interspeech,
  title     = {{Enhancing Multilingual Voice Toxicity Detection with Speech-Text Alignment}},
  author    = {Joseph Liu and Mahesh Kumar Nandwana and Janne Pylkk{"o}nen and Hannes Heikinheimo and Morgan McGuire},
  year      = {2024},
  booktitle = {{Interspeech 2024}},
  pages     = {4298--4302},
  doi       = {10.21437/Interspeech.2024-1228}
}

@inproceedings{ranjan25b_interspeech,
  title     = {{Multimodal Zero-Shot Framework for Deepfake Hate Speech Detection in Low-Resource Languages}},
  author    = {Rishabh Ranjan and Likhith Ayinala and Mayank Vatsa and Richa Singh},
  year      = {2025},
  booktitle = {{Interspeech 2025}},
  pages     = {1678--1682},
  doi       = {10.21437/Interspeech.2025-2668}
}

@inproceedings{luo25_interspeech,
  title     = {{ToxicTone: A Mandarin Audio Dataset Annotated for Toxicity and Toxic Utterance Tonality}},
  author    = {Yu-Xiang Luo and Yi-Cheng Lin and Ming-To Chuang and Jia-Hung Chen and I-Ning Tsai and Pei Xing Kiew and Yueh-Hsuan Huang and Chien-Feng Liu and Yu-Chen Chen and Bo-Han Feng and Wenze Ren and Hung-yi Lee},
  year      = {2025},
  booktitle = {{Interspeech 2025}},
  pages     = {4008--4012},
  doi       = {10.21437/Interspeech.2025-679}
}

@incollection{sakhare2023deep,
  title={Deep learning-based intelligent systems for audio abuse prediction: A survey},
  author={Sakhare, Kaustubh V and Kulkarni, Radhika V},
  booktitle={AI-Based Metaheuristics for Information Security and Digital Media},
  pages={35--47},
  year={2023},
  publisher={Chapman and Hall/CRC}
}

@article{Garg2022Handling,
title={Handling Bias in Toxic Speech Detection: A Survey},
author={Tanmay Garg and Sarah Masud and Tharun Suresh and Tanmoy Chakraborty},
journal={ACM Computing Surveys},
year={2022},
volume={55},
pages={1 - 32},
doi={10.1145/3580494}
}

@article{Tundik2019Assessing,
title={Assessing the Semantic Space Bias Caused by ASR Error Propagation and its Effect on Spoken Document Summarization},
author={Máté Ákos Tündik and Valér Kaszás and György Szaszák},
year={2019},
pages={1333-1337},
doi={10.21437/interspeech.2019-2154}
}

@article{Wang2025CLIP6D,
title={CLIP-6D: Empowering CLIP as a Zero-Shot 6D Pose Estimator Through Generalizable Object-Specific Representations},
author={Hua Wang and Hong Liu and Jiale Ren and Mingxin Tan and Zhongzien Jiang},
journal={Proceedings of the 33rd ACM International Conference on Multimedia},
year={2025},
doi={10.1145/3746027.3754497}
}

@article{Qian2024Online,
title={Online Zero-Shot Classification with CLIP},
author={Qi Qian and Juhua Hu},
year={2024},
pages={462-477},
doi={10.48550/arxiv.2408.13320}
}

@inproceedings{elizalde2023clap,
  title={Clap learning audio concepts from natural language supervision},
  author={Elizalde, Benjamin and Deshmukh, Soham and Al Ismail, Mahmoud and Wang, Huaming},
  booktitle={ICASSP 2023-2023 IEEE International Conference on Acoustics, Speech and Signal Processing (ICASSP)},
  pages={1--5},
  year={2023},
  organization={IEEE}
}

@article{Chhabra2023A,
title={A literature survey on multimodal and multilingual automatic hate speech identification},
author={Anusha Chhabra and D. Vishwakarma},
journal={Multimedia Systems},
year={2023},
pages={1-28},
doi={10.1007/s00530-023-01051-8}
}

@misc{radford2021clip,
      title={Learning Transferable Visual Models From Natural Language Supervision}, 
      author={Alec Radford and Jong Wook Kim and Chris Hallacy and Aditya Ramesh and Gabriel Goh and Sandhini Agarwal and Girish Sastry and Amanda Askell and Pamela Mishkin and Jack Clark and Gretchen Krueger and Ilya Sutskever},
      year={2021},
      eprint={2103.00020},
      archivePrefix={arXiv},
      primaryClass={cs.CV},
      url={https://arxiv.org/abs/2103.00020}, 
}

@misc{oord2019infoceloss,
      title={Representation Learning with Contrastive Predictive Coding}, 
      author={Aaron van den Oord and Yazhe Li and Oriol Vinyals},
      year={2019},
      eprint={1807.03748},
      archivePrefix={arXiv},
      primaryClass={cs.LG},
      url={https://arxiv.org/abs/1807.03748}, 
}

@article{khosla2020supervised,
  title={Supervised contrastive learning},
  author={Khosla, Prannay and Teterwak, Piotr and Wang, Chen and Sarna, Aaron and Tian, Yonglong and Isola, Phillip and Maschinot, Aaron and Liu, Ce and Krishnan, Dilip},
  journal={Advances in neural information processing systems},
  volume={33},
  pages={18661--18673},
  year={2020}
}

@article{Kim2022Integrated,
title={Integrated Parameter-Efficient Tuning for General-Purpose Audio Models},
author={Ju-Ho Kim and Ju-Sung Heo and Hyun-Seo Shin and Chanmann Lim and Ha-jin Yu},
journal={ArXiv},
year={2022},
volume={abs/2211.02227},
doi={10.48550/arxiv.2211.02227}
}

@article{Liang2024AAT,
title={AAT: Adapting Audio Transformer for Various Acoustics Recognition Tasks},
author={Yun Liang and Hai Lin and Shaojian Qiu and Yihang Zhang},
journal={ICASSP 2024 - 2024 IEEE International Conference on Acoustics, Speech and Signal Processing (ICASSP)},
year={2024},
pages={1361-1365},
doi={10.1109/icassp48485.2024.10447544}
}

@article{Unanue2023T3L,
title={T3L: Translate-and-Test Transfer Learning for Cross-Lingual Text Classification},
author={Inigo Jauregi Unanue and Gholamreza Haffari and M. Piccardi},
journal={Transactions of the Association for Computational Linguistics},
year={2023},
volume={11},
pages={1147-1161},
doi={10.1162/tacl_a_00593}
}

@article{Raza2025A,
title={A Machine Learning Approach of Text Classification for High‐ and Low‐Resource Languages},
author={Muhammad Owais Raza and N. Mahoto and A. Shaikh and Nazia Pathan and Hani Alshahrani and M. A. Elmagzoub},
journal={Computational Intelligence},
year={2025},
volume={41},
doi={10.1111/coin.70114}
}

@misc{radford2022whisper,
  doi = {10.48550/ARXIV.2212.04356},
  url = {https://arxiv.org/abs/2212.04356},
  author = {Radford, Alec and Kim, Jong Wook and Xu, Tao and Brockman, Greg and McLeavey, Christine and Sutskever, Ilya},
  title = {Robust Speech Recognition via Large-Scale Weak Supervision},
  publisher = {arXiv},
  year = {2022},
  copyright = {arXiv.org perpetual, non-exclusive license}
}

\appendix
\renewcommand{\arraystretch}{1.2}

\section{Appendix}

\subsection{Additional Cross-Lingual and LOLO Score Tables}
\label{app:extra_tabs}

Tables~\ref{tab:variance}--\ref{tab:lolo_task_b_appendix} give the full per-language, per-shot values underlying the analyses in Sections~\ref{sec:variance} and~\ref{sec:lolo_discussion} (mean over 5 support-set draws for $k\geq1$; single run at $k=0$; accuracy not tracked per-seed). The corresponding cross-lingual comparison against ADIMA is given in Table~\ref{tab:crosslingual_mf1_main}.

{\renewcommand{\arraystretch}{1.2}
\begin{table*}[t]
\centering
\small
\begin{tabular}{@{}lcccccc@{}}
\toprule
Language  & 1-shot        & 5-shot        & 10-shot       & 25-shot       & 50-shot       & 0-shot \\
\midrule
Bengali   & 74.8 $\pm$0.3 & 75.0 $\pm$2.3 & 74.5 $\pm$0.9 & 74.8 $\pm$1.3 & 74.2 $\pm$1.3 & 76.3 \\
Bhojpuri  & 70.0 $\pm$0.9 & 70.5 $\pm$0.7 & 69.9 $\pm$1.4 & 70.2 $\pm$0.8 & 69.3 $\pm$1.0 & 68.9 \\
Gujarati  & 74.1 $\pm$1.0 & 73.7 $\pm$0.9 & 74.6 $\pm$0.8 & 73.7 $\pm$1.6 & 73.6 $\pm$0.7 & 74.6 \\
Haryanvi  & 77.0 $\pm$1.2 & 77.2 $\pm$1.8 & 77.6 $\pm$0.8 & 76.8 $\pm$0.6 & 77.4 $\pm$1.0 & 77.4 \\
Hindi     & 77.2 $\pm$0.3 & 75.6 $\pm$0.9 & 76.4 $\pm$1.0 & 77.3 $\pm$1.1 & 76.8 $\pm$0.8 & 77.8 \\
Kannada   & 75.6 $\pm$0.4 & 75.7 $\pm$0.6 & 76.0 $\pm$1.4 & 75.6 $\pm$0.3 & 74.7 $\pm$1.0 & 76.8 \\
Malayalam & 76.9 $\pm$1.0 & 77.9 $\pm$1.0 & 77.4 $\pm$0.6 & 76.9 $\pm$1.6 & 76.3 $\pm$0.5 & 76.9 \\
Odia      & 79.4 $\pm$1.1 & 78.2 $\pm$0.6 & 78.1 $\pm$1.1 & 78.4 $\pm$1.2 & 78.8 $\pm$1.2 & 81.5 \\
Punjabi   & 81.7 $\pm$0.6 & 82.2 $\pm$0.9 & 82.0 $\pm$0.7 & 81.8 $\pm$1.3 & 82.2 $\pm$0.6 & 82.3 \\
Tamil     & 72.0 $\pm$1.1 & 71.9 $\pm$0.4 & 72.4 $\pm$1.1 & 70.9 $\pm$1.8 & 72.5 $\pm$0.9 & 73.0 \\
\bottomrule
\end{tabular}
\caption{Cross-lingual macro-F1 (mean $\pm$ std over 5 seeds), projection-only. $k{=}0$ is a single deterministic run (no support-set sampling).}
\label{tab:variance}
\end{table*}
} % end \renewcommand{\arraystretch}

{\renewcommand{\arraystretch}{1.2}
\begin{table}[t]
\centering
\small
\begin{tabular}{@{}lcc@{}}
\toprule
Language & Best shot (proj-only, mean) & 0-shot optimal? \\
\midrule
Bengali   & 0  & yes \\
Bhojpuri  & 5  & no  \\
Gujarati  & 0  & yes \\
Haryanvi  & 10 & no$^\dagger$ \\
Hindi     & 0  & yes \\
Kannada   & 0  & yes \\
Malayalam & 5  & no$^\ddagger$ \\
Odia      & 0  & yes \\
Punjabi   & 0  & yes \\
Tamil     & 0  & yes \\
\bottomrule
\end{tabular}
\caption{Best-performing shot size per language (cross-lingual, projection-only; Table~\ref{tab:variance}). $\dagger$: gain within one std. $\ddagger$: gain approximately one std.}
\label{tab:shot-optimal}
\end{table}
} % end \renewcommand{\arraystretch}

{\renewcommand{\arraystretch}{1.2}
\begin{table}[t]
\centering
\small
\begin{tabular}{@{}lccc@{}}
\toprule
Prompt variant & mean MF1 & min & max \\
\midrule
P1\_basic      & 53.6 & 45.1 & 60.2 \\
P2\_hate       & 65.6 & 62.3 & 73.6 \\
P3\_offensive  & 65.2 & 59.8 & 72.6 \\
P4\_aggressive & 64.3 & 58.4 & 75.1 \\
\bottomrule
\end{tabular}
\caption{Zero-shot prompt sensitivity (macro-F1 across the ten languages); min/max taken across languages per prompt variant.}
\label{tab:prompt-sensitivity}
\end{table}
} % end \renewcommand{\arraystretch}

%======================
% Table A1: LOLO projection-only
%======================
\begin{table}[t]
\centering
\small
\setlength{\tabcolsep}{5pt}
\begin{tabular}{lcc}
\toprule
Language & Macro-F1 (mean) & Best shot \\
\midrule
Bengali    & 76.3 & 0-shot  \\
Bhojpuri   & 70.6 & 5-shot  \\
Gujarati   & 74.6 & 0-shot  \\
Haryanvi   & 77.7 & 10-shot \\
Hindi      & 77.8 & 0-shot  \\
Kannada    & 76.8 & 0-shot  \\
Malayalam  & 77.5 & 10-shot \\
Odia       & 81.5 & 0-shot  \\
Punjabi    & 82.3 & 0-shot  \\
Tamil      & 73.0 & 0-shot  \\
\bottomrule
\end{tabular}
\caption{Best LOLO results, projection-only (mean macro-F1 over 5 draws, $k\geq1$; single run at $k=0$).}
\label{tab:lolo_task_a_appendix}
\end{table}

%======================
% Table A2: LOLO projection+fine-tuning
%======================
\begin{table}[t]
\centering
\small
\setlength{\tabcolsep}{5pt}
\begin{tabular}{lcc}
\toprule
Language & Macro-F1 (mean) & Best shot \\
\midrule
Bengali    & 76.3 & 0-shot  \\
Bhojpuri   & 70.2 & 10-shot \\
Gujarati   & 74.6 & 0-shot  \\
Haryanvi   & 77.4 & 0-shot  \\
Hindi      & 77.8 & 0-shot  \\
Kannada    & 76.8 & 0-shot  \\
Malayalam  & 76.9 & 0-shot  \\
Odia       & 81.5 & 0-shot  \\
Punjabi    & 82.3 & 0-shot  \\
Tamil      & 73.0 & 0-shot  \\
\bottomrule
\end{tabular}
\caption{Best LOLO results, projection+fine-tuning (mean macro-F1 over 5 draws, $k\geq1$; single run at $k=0$).}
\label{tab:lolo_task_b_appendix}
\end{table}

\begin{figure*}[h]
    \centering
    \includegraphics[width=0.62\linewidth]{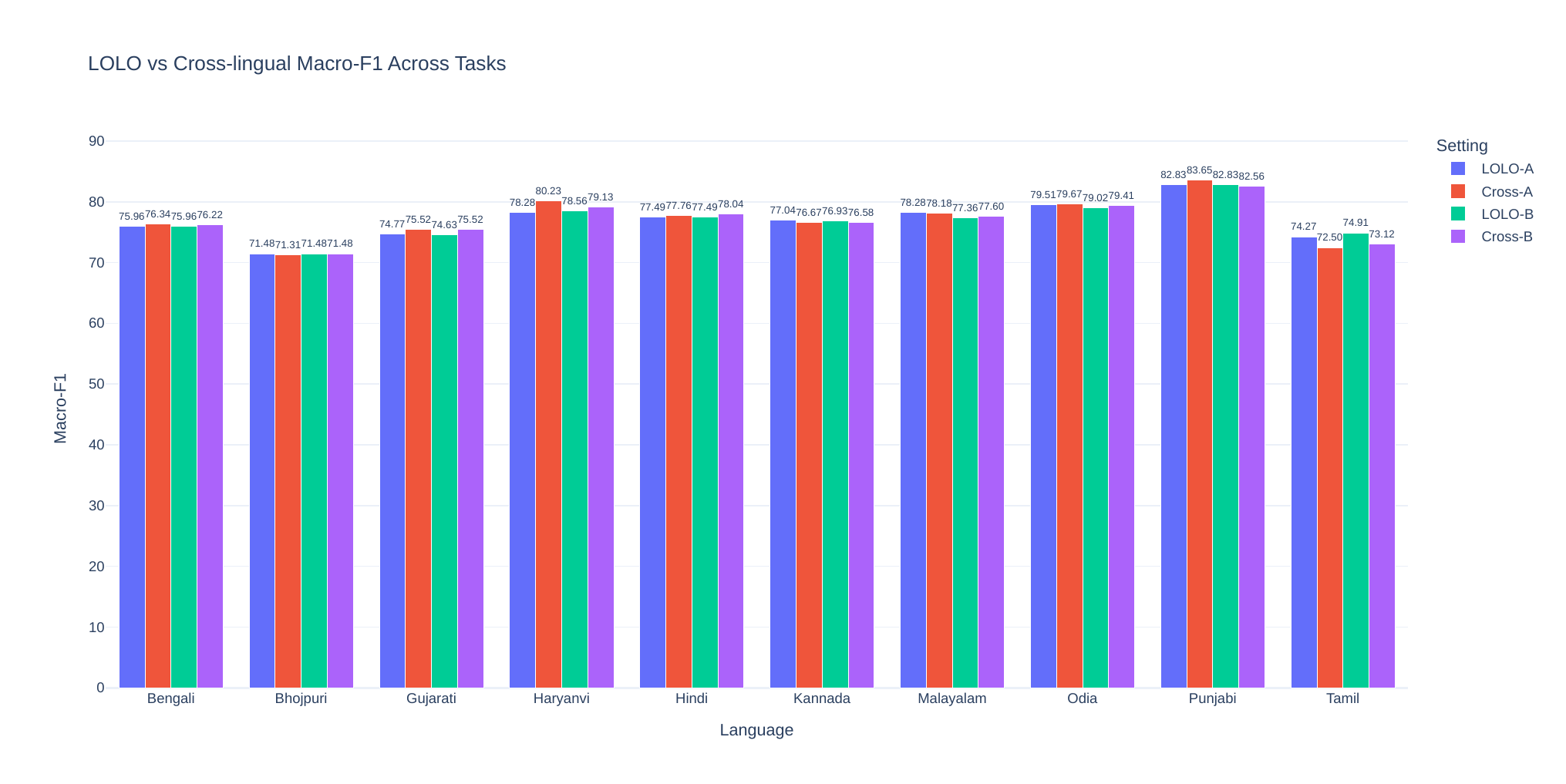}
    \caption{Language-wise LOLO vs.\ cross-lingual macro-F1, projection-only and projection+fine-tuning.}
    \label{fig:app_lolo_cross_macro}
\end{figure*}

\subsection{Hyper-parameters and Training Details}
\label{app:hyperparams}

All experiments use \texttt{laion/clap-htsat-unfused}~\footnote{https://hf.co/laion/clap-htsat-unfused} as the base CLAP encoder. Audio is resampled and padded/truncated to a fixed duration before feature extraction. Support sets use $k\in\{0,1,5,10,25,50\}$ examples per class per language ($k{=}0$: frozen embeddings, no adaptation). Training runs for 50 epochs, learning rate 1e-4, AdamW with default warm-up, on the Apple MPS backend (PyTorch). Projection-only adaptation (Section~\ref{sec:method}) uses a temperature-scaled supervised contrastive loss~\citep{khosla2020supervised}; projection+fine-tuning additionally updates the final encoder blocks with a smaller encoder learning rate.
Downstream classification uses an SVM and a scikit-learn ANN-MLP (Section~\ref{sec:method}); the higher-macro-F1 model is reported per language, with accuracy as a tie-breaker. Random seeds are fixed for support-set sampling, adaptation, and classifier training throughout.

\subsection{Additional Cross-Lingual and LOLO Visualisations}\label{app:extra_plots}

Figure~\ref{fig:app_lolo_cross_macro} visualizes the per-language LOLO vs.\ cross-lingual macro-F1 gap discussed in Section~\ref{sec:lolo_discussion}.

\end{document}